# Transforming Telemedicine Through Big Data Analytics

Michael Coakley, Giancarlo Crocetti, Phil Dressner, Wanda Kellum, Tamba Lamin
Seidenberg School of Computer Science and Information Systems
Pace University
White Plains, NY, US
{mc77168w, gc47601w, pd50340n, wk59882w, tl98810w}@pace.edu

*Abstract*— A look at how big data is transforming telemedicine to provide better care by tapping into a larger source of patient information. Telemedicine will have a profound impact on patient care, increase access and quality, and represent an opportunity to keep health care costs down. Data generated by smart devices will enable the real-time monitoring of chronic diseases, allowing optimal dosage of drugs and improve patient outcomes.

## I. INTRODUCTION

Telemedicine refers to the use of modern telecommunication capabilities for medical information exchange between a care provider and patients in order to improve health outcome. Started out over forty years ago with the use of telecommunication technologies to extend hospital care to patients located in remote areas, telemedicine has spread rapidly and is now becoming integrated into the daily operations of hospital, specialty departments, home health agencies, and private physicians [1].

Telehealth solutions, promise to have a profound impact on patient care, its quality and safety, and can also help drive costs out of the healthcare system. For instance, enabling the chronically ill and elderly to receive care from home reduces the number of hospital admissions and readmissions, which are riddled with expense and risk of exposure to other illnesses.

## II. TELEMEDICNE AS A SERVICE

Telemedicine is primarily an enabling technology which provides primary care and patient monitoring.

Primary care and specialist referral services may involve a primary care or allied health professional providing a consultation with a patient or a specialist assisting the primary care physician in rendering a diagnosis. This may involve the use of live interactive video or the use of store and forward transmission of diagnostic images, vital signs and/or video clips along with patient data for later review.

Remote patient monitoring, including home telehealth, user devices to remotely collect and send data to a home health agency or a remote diagnostic testing facility (RDTF) for interpretation. Such applications might include a specific vital sign, such as blood glucose or heart ECG or a variety of indicators for homebound patients. Such services can be used to supplement the use of visiting nurses.

Consumer medical and health information includes the use of the Internet and wireless devices for consumers to obtain specialized health information and on-line discussion groups to provide peer-to-peer support [1].

The ubiquitous presence of mobile infrastructure even in developing country translates into improved access to health for people with limited interaction to health providers, filling an important gap in bringing health care into underserved and underprivileged communities. The easy access to telecommunication infrastructure also allows the possibility of training local health professionals through special medical education seminars remotely.

## III. BIG DATA IN TELEMEDICINE

From wearable devices to nanotechnologies to self-tests, our life will soon be filled with a plethora of devices that will constantly monitor our health. These devices will constantly generate data that will be collected and analyzed real time.

Instead of sampling the data for analysis, systems will detect disease patterns on-the-fly and alert the patient and the doctor at the first sign of anomaly, which is particularly important for chronic illness like atrial fibrillation or blood clotting to name just a few [2].

The variability of this data will be so great that we will be able to detect differences on how the same disease affects different sub-population and administer the proper medication according to the proper phenotype. Through the use of wireless and nanotechnologies we will be able to administer the right drug, at the right level, at the right time by virtually eliminate side effects. This will be possible thanks to personalize medicine which is collecting a huge amount of genomic data to understand how drugs are metabolized differently by different people. [3]

The research team at the Mt. Sinai Hospital in New York, for example, is linking together the sequencing information of patients suffering of various forms of cancer and already identified genomic pathways for the development of novel therapeutic and diagnostic approaches for human diseases through integration and analysis of molecular and clinical data, and also perform research to develop and evaluate methods to incorporate genomic sequencing data into clinical practice [4].

More data allows the research and development of new and improved methods for the diagnosis, prevention, and treatment or rare and common genetic diseases.

## IV. WHY BIG DATA IS IMPORTANT

Big data usually includes data sets from diverse sources and with sizes well beyond the ability of commonly used software tools to ingest, manage, and process the data within an acceptable amount of time. The bad news, however, do not stop here: things will only get worst with data to grow 50 folds by 2020 [5].

Many industries are following this issue very closely, with some companies like Amazon, Google, and Netflix taking the lead in attacking the problem heads on.

For telemedicine this means the necessity, for the medical support staff, to extract the necessary data for patients to pass along to medical professional. The ability of processing such huge amount of information will translate into improved medical knowledge that benefits research and patient alike.

Big data represents a revolution in medical research and the time when we struggled to obtain samples in order to study a particular population are over. Today, with the availability of large and diverse datasets coming from many parts of the world allows us to consider the size of our sample as *n=All*: this will force a complete overhaul of our research methods which are still based on century-old sampling techniques [6].